\documentclass[aps,11pt,pra,superscriptaddress,preprint]{revtex4}

\usepackage[T1]{fontenc} 
\usepackage{slashed}
\usepackage[utf8x]{inputenc}
\usepackage{color}
\usepackage[normalem]{ulem}
\usepackage{bm}
\usepackage{graphicx}

\begin{document}
\title{\textbf{Effective field theories: from Cosmology to Quark Gluon Plasma}
}
\author{
Simone Biondini   \\
{\em Technische Universit\"at M\"unchen}, 
\\
{\em Physik Department I, Garching 85747, Germany  } \\
}

\preprint{TUM-EFT 49/14}
\begin{abstract}
Cosmology and particle physics come across a tight connection in the attempt to reproduce and understand quantitatively the results of experimental findings. Indeed, the quark gluon plasma (QGP) found at colliders and the baryon asymmetry provided by the WMAP collaboration are examples where to apply field theoretical techniques in issues relevant for Cosmology. 

In the simplest leptogenesis framework, heavy Majorana neutrinos are at the origin of the baryon asymmetry. The non-relativistic regime appears to be relevant during the lepton asymmetry generation where the interactions among particles occur in a thermal medium.   

We discuss the development of an effective field theory (EFT) for non-relativistic Majorana particles to address calculations at finite temperature. We show an application of such a method to the case of a heavy Majorana neutrino decaying in a hot and dense plasma of Standard Model (SM) particles.

These techniques are analogous to those widely used for the investigation of heavy-ion collisions at colliders by exploiting hard probes. Finally we sketch some commonalities between Majorana neutrinos and bound state of heavy quarks in medium.
\end{abstract}
\maketitle
\section{Introduction}
The physics of the early universe represents a challenge both theoretically and experimentally. Indeed, during the first stages of the universe evolution, particles interact at very high energies, density and temperatures. At present, such extreme conditions might be reproduced only in dedicated facilities.

We observed recently hints at the ALICE experiment \cite{Aamodt:2010jd}, and previously at RICH \cite{RICH}, that the quark gluon plasma is possibly formed. This is a phase in which quarks and gluons are no longer confined in hadrons and this transition has likely occurred during the universe evolution. Confirming the existence of this phase of matter and understanding its properties might shed light on our comprehension about particles interactions in the early universe. Quarkonia suppression in medium has been considered a promising probe to study QGP in heavy-ion collisions \cite{Matsui:1986dk}.  

Other important processes taking place in the hot and dense universe are the dark matter and baryon asymmetry generation. The WMAP collaboration provides accurate values for the dark matter abundance and the amount of baryon asymmetry \cite{Hinshaw:2012aka}. At variance with the quark gluon plasma that is explained by quantum cromodynamics, some physics beyond the SM is needed to account for such evidences. Let us focus on the generation of the baryon asymmetry and on a possible attractive solution, embedded in the class of models called leptogenesis. 

In its simplest scenario, a net lepton asymmetry is produced in the CP violating decays of heavy Majorana neutrinos into SM leptons and anti-leptons in different amounts \cite{Fukugita:1986hr}. The lepton asymmetry is then partially reprocessed in a baryon asymmetry via the sphalerons transitions. The interactions of heavy neutrinos with particles in the plasma occur in a hot medium and one has to take into account a field theoretical approach that includes thermal effects. 

Either in the case of heavy quarkonium in a quark gluon plasma or heavy Majorana neutrinos in the early universe,  the presence of a thermal medium  may induce some effects. Much progress has been done in this direction and we discuss here the EFT approach to address the issue. 

In section \ref{First section}, we introduce the EFT formalism we are going to use for a non-relativistic particle. We show how to reproduce the neutrino thermal width by using the EFT techniques in section \ref{Second section}, whereas we sketch the case of heavy quarkonium in a weakly interacting QGP in section \ref{Third section}.     

\section{Effective Lagrangian for non-relativistic particles}
\label{First section}

We define a particle to be heavy in this contest when its mass, $M$, is bigger than any other scale that characterizes the medium. For a system in equilibrium the main thermodynamical scale is the temperature, $T$. The situation we want to consider is $M \gg T$ and in such a case the heavy particle is also non-relativistic. Hence, we can devise an EFT Lagrangian that properly describes the low-energy modes of the heavy particle as the relevant degrees of freedom, labelled with the field $N$ in the following. The effective Lagrangian has the general form
\begin{equation}
\mathcal{L}_{{\rm{EFT}}}=N^{\dagger} i D_{0} N + \sum_{n} \, c_{n}\left( \frac{\mu}{M}\right) \frac{\mathcal{O}_{n}(\mu,T)}{M^{d_{n}-4}} + \mathcal{L}_{light}  \, .
\label{Lag0}
\end{equation}
The heavy particle is considered in a reference frame at rest up to momentum fluctuations much smaller than $M$. 
The heavy particle sector is organized as an expansion in $1/M$. The effective operators, describing the interaction among the heavy and light degrees of freedom, are suppressed with the proper power of $M$ in order to keep the Lagrangian density of dimension four. The higher the operator dimension the bigger the suppression. The light degrees of freedom are embedded in $\mathcal{L}_{light}$. 

The Wilson coefficients, $c_n$, are the parameters of the low-energy theory and one has to compute them by the matching procedure. Since they encode the effects of the high energy scale, $M$, one may ignore and set to zero any other scale in the matching calculation, namely $T=0$. This turns out to be a helpful simplification in the case of calculations in a thermal bath. Indeed, the Wilson coefficients are obtained by matching  matrix elements at energy scales of order $\Lambda$ that satisfies $M \gg \Lambda \gg T$. The temperature only affects the calculation of observables in the low-energy Lagrangian, as we are going to show in the case of the neutrino thermal width in the next section.      

\begin{figure}[h]
\centering
\includegraphics[scale=0.45]{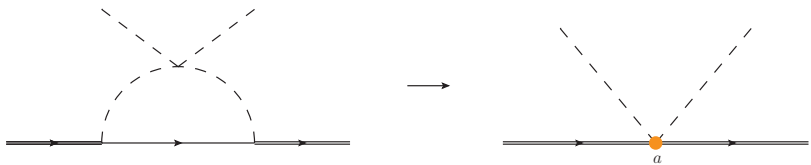}
\caption{\label{Fig0} Relevant diagram for the matching computation of the dimension five operator. The solid double line stands for the non-relativistic Majorana neutrino, the dashed line stands for the Higgs boson and the solid line for the SM leptons.}
\end{figure}

\section{Thermal width for heavy Majorana neutrinos}
\label{Second section}
We are interested in the thermal width of a Majorana neutrino induced by a thermal bath. We start with the fundamental Lagrangian that includes one additional neutrino to the SM content. The additional lepton is also called sterile or right-handed neutrino, it has a Majorana mass and interacts only via Yukawa couplings with SM Higgs and leptons. The Lagrangian reads
\begin{equation}
\mathcal{L}=\mathcal{L}_{\rm{SM}} + \frac{1}{2} \bar{\psi} i \slashed{\partial}  \psi  - \frac{M}{2} \bar{\psi}\psi - F_{f}\bar{L}_{f} \tilde{\phi} P_{R}\psi  - F^{*}_{f}\bar{\psi}P_{L} \tilde{\phi}^{\dagger}  L_{f} \, ,
\label{Lag1}
\end{equation}
where $\mathcal{L}_{\rm{SM}}$ is the SM Lagrangian, $\psi$ is the Majorana neutrino field, $F_f$ is the Yukawa coupling, $L_f$ stands for the SM lepton doublet with flavour $f$, $\tilde{\phi}=i\sigma_2\phi^{*}$ where $\phi$ is the Higgs field and $P_{L(R)}$ is the left (right) chiral projector. This Lagrangian is not enough to have leptogenesis because  there is only one sterile neutrino. However, we take the Lagrangian in (\ref{Lag1}) as a toy model to implement the EFT treatment for the thermal width (in turn connected with the thermal production rate \cite{Laine:2011pq}). We consider the following hierarchy of scales $M \gg T \gg M_W$, where $M_W$ sets the electroweak scale according to the standard thermal leptogenesis scenario. 

The first step consists in integrating out the energy modes of order $M$ so that we are left with non-relativistic excitations of Majorana neutrinos. The low-energy Lagrangian reads, according to the prototype in eq. (\ref{Lag0}), as follows
\begin{equation}
\mathcal{L}_{{\rm{EFT}}}=\mathcal{L}_{\rm{SM}} + N^{\dagger}\left( i \partial_{0} -i \frac{\Gamma_{0}}{2}\right)N + \frac{\mathcal{L}^{(1)}}{M} + \frac{\mathcal{L}^{(2)}}{M^2} + \frac{\mathcal{L}^{(3)}}{M^3} + \mathcal{O} \left( \frac{1}{M^4}\right) \, , 
\end{equation} 
where $\Gamma_0$ is the decay with at $T=0$. $\mathcal{L}^{(1)}$, $\mathcal{L}^{(2)}$ and $\mathcal{L}^{(3)}$ contain respectively dimension five, six and seven operators. We focus on the dimension five operator in  $\mathcal{L}^{(1)}$, $a \, N^{\dagger}N \phi^{\dagger}\phi$, in order to show our procedure. Indeed, symmetry arguments allow for only one dimension five operator that describes the effective interaction between a non-relativistic Majorana neutrino and a SM Higgs boson. By integrating out the energy modes of order $M$, the one loop process in the fundamental theory is matched onto a four-particle interaction, as shown in Figure \ref{Fig0}. Indeed, at energies much smaller than $M$, the vertices in the high energy theory can not be resolved. We stress that standard $T=0$ techniques are exploited in the matching calculation, because the temperature is such that $M \gg \Lambda \gg T$, where $\Lambda$ is the ultraviolet cut-off in the low-energy theory. Therefore $T$ is set to zero and it does not enter the calculation. The matching coefficient reads ${\rm{Im}}(a)= -(3\lambda |F_f|^2 )/(8 \pi)$ and determines the low-energy neutrino-Higgs interaction. We extracted only the imaginary part since it is the one relevant for the thermal width. 

\begin{figure}[h]
\centering
\includegraphics[scale=0.5]{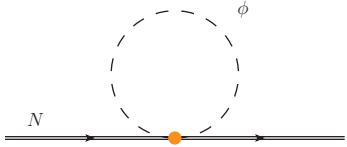}
\caption{\label{Fig1} The tadpole diagram in the low-energy theory is shown.}
\end{figure}

Once the effective Lagrangian is determined, one can calculate the thermal corrections to the width. This is carried out in the low-energy theory, where the dynamical energy modes are those of order $T$. In the present formalism, the thermal width is obtained by computing the tadpole diagram in Figure \ref{Fig1}. A Higgs boson from the thermal bath runs in the loop diagram and the leading thermal correction reads \cite{Laine:2011pq}, \cite{Salvio:2011sf}
\begin{equation}
\Gamma_{a}=2\frac{{\rm{Im}}(a)}{M} \langle \phi^{\dagger}(0)\phi(0) \rangle \, =-\lambda\frac{ |F_f|^2 M}{8 \pi}\left( \frac{T}{M} \right)^2 \, . 
\end{equation}   
We notice that the derivation of the thermal width is divided into two steps: a one loop calculation in vacuum (matching), and a one loop computation in the EFT (thermal tadpole). In a full relativistic thermal field theory derivation for the thermal width such a simplification can barely arise, whereas it is typical of the EFT approach.    
\section{Quarkonium suppression in QGP and conclusions}
\label{Third section}
As we mentioned in the introduction, quarkonia suppression in medium may be useful to probe the QGP features. The main difference with respect to Majorana neutrinos is that heavy quarkonium is a bound state of heavy quarks ($Q\bar{Q}$).

This complicates much more the treatment in a thermal medium because of the interplay between non-relativistic and thermodynamic scales. The former are $m$, $m v$, and $mv^2$, respectively the mass of the heavy quark, the inverse size of the bound state and the binding energy. The temperature, $T$, and the Debye mass, $m_D \sim gT$, are the thermodynamic ones. An EFT treatment for a wide range of scales configurations has been established in \cite{Brambilla:2008cx}. The main outcome is a rigorous derivation of the static potential that comprises a real and imaginary part with thermal corrections. In particular, the imaginary part of the potential is related with the thermal width of the quarkonium that can be traced back to the interactions between the bound state and the partons in the QCD medium. This is in complete analogy with the Majorana neutrino thermal width that arises from the interactions with the SM particles in the thermal bath. Therefore, the developments of resummation techniques in hot QCD and Cosmology may benefit both fields.

\section{Acknowledgments}
I would like to thank N. Brambilla, M. Escobedo and A. Vairo for the collaboration on the work presented here based on \cite{Biondini:2013xua}, and the organizers of the \textit{4th Young Researchers Workshop} for the possibility to give a talk. 


%

\begin{thebibliography}{99}

\bibitem{Aamodt:2010jd}
  K.~Aamodt {\it et al.}  [ALICE Collaboration],
  Phys.\ Lett.\ B {\bf 696} (2011) 30
  [arXiv:1012.1004 [nucl-ex]].
\bibitem{RICH}
STAR collaboration,  Phys.\ Lett.\ B {\bf 567} (2003) 167.

\bibitem{Matsui:1986dk}
  T.~Matsui and H.~Satz,
  Phys.\ Lett.\ B {\bf 178} (1986) 416.
  
  
  \bibitem{Hinshaw:2012aka}
  G.~Hinshaw {\it et al.}  [WMAP Collaboration],
  Astrophys.\ J.\ Suppl.\  {\bf 208} (2013) 19
  [arXiv:1212.5226 [astro-ph.CO]].
  
  
\bibitem{Fukugita:1986hr}
  M.~Fukugita and T.~Yanagida,
  Phys.\ Lett.\ B {\bf 174} (1986) 45.
  
 

  
  \bibitem{Laine:2011pq}
  M.~Laine and Y.~Schroder,
  JHEP {\bf 1202} (2012) 068
  [arXiv:1112.1205 [hep-ph]].
  
  \bibitem{Salvio:2011sf}
  A.~Salvio, P.~Lodone and A.~Strumia,
  JHEP {\bf 1108} (2011) 116
  [arXiv:1106.2814 [hep-ph]].
  
  \bibitem{Brambilla:2008cx}
  N.~Brambilla, J.~Ghiglieri, A.~Vairo and P.~Petreczky,
  Phys.\ Rev.\ D {\bf 78} (2008) 014017
  [arXiv:0804.0993 [hep-ph]].
  
  \bibitem{Biondini:2013xua}
  S.~Biondini, N.~Brambilla, M.~A.~Escobedo and A.~Vairo,
  JHEP {\bf 1312} (2013) 028
  [arXiv:1307.7680, arXiv:1307.7680].
\end{thebibliography}
\end{document}